\documentclass[conference]{IEEEtran}
\IEEEoverridecommandlockouts
\usepackage{cite}
\usepackage{amsmath,amssymb,amsfonts}
\usepackage{algorithmic}
\usepackage{balance}
\usepackage{graphicx}
\usepackage{textcomp}
\usepackage{xcolor}
\def\BibTeX{{\rm B\kern-.05em{\sc i\kern-.025em b}\kern-.08em
    T\kern-.1667em\lower.7ex\hbox{E}\kern-.125emX}}
\begin{document}

\title{Multi-Partner Project: COIN-3D -- Collaborative Innovation in 3D VLSI Reliability }

\author{\IEEEauthorblockN{~George~Rafael~Gourdoumanis\textsuperscript{†},~Fotoini~Oikonomou\textsuperscript{†},~Maria~Pantazi-Kypraiou\textsuperscript{†},~Pavlos~Stoikos\textsuperscript{†},\\~Olympia~Axelou\textsuperscript{†},~Athanasios~Tziouvaras\textsuperscript{†}, ~Georgios~Karakonstantis\textsuperscript{†},~Tahani~Aladwani\textsuperscript{+},\\~Christos~Anagnostopoulos\textsuperscript{+},~Yixian~Shen\textsuperscript{\ddag},~Anuj~Pathania\textsuperscript{\ddag},~Alberto~Garcia-Ortiz\textsuperscript{*}, and~George~Floros\textsuperscript{†\S}}

\IEEEauthorblockA{\textsuperscript{†}Department of Electrical and Computer Engineering, 
University of Thessaly, Greece \\ 
\{ggeorgios-r, fotoikon, mpantazi-, pastoikos, oaxelou, attziouv, gkarakon, gefloros\}@e-ce.uth.gr\\
\textsuperscript{+}School of Computing Science, University of Glasgow, UK\\
\{tahani.aladwani, christos.anagnostopoulos\}@glasgow.ac.uk\\
{\ddag}University of Amsterdam, The Netherlands \\
\{y.shen,~a.pathania\}@uva.nl\\
\textsuperscript{*}Institute of Electrodynamics and Microelectronics, University of Bremen, Germany \\
agarcia@item.uni-bremen.de\\
\textsuperscript{\S}Department of Electronic and Electrical Engineering, Trinity College Dublin, Ireland\\florosg@tcd.ie}
}
\maketitle

\begin{abstract}
As semiconductor manufacturing advances from the 3-nm process toward the sub-nanometer regime and transitions from FinFETs to gate-all-around field-effect transistors (GAAFETs), the resulting complexity and manufacturing challenges continue to increase. In this context, 3D chiplet-based approaches have emerged as key enablers to address these limitations while exploiting the expanded design space. Specifically, chiplets help address the lower yields typically associated with large monolithic designs. This paradigm enables the modular design of heterogeneous systems consisting of multiple chiplets (e.g., CPUs, GPUs, memory) fabricated using different technology nodes and processes. Consequently, it offers a capable and cost-effective strategy for designing heterogeneous systems.

This paper introduces the Horizon Europe Twinning project COIN-3D (Collaborative Innovation in 3D VLSI Reliability), which aims to strengthen research excellence in 2.5D/3D VLSI systems reliability through collaboration between leading European institutions. More specifically, our primary scientific goal is the provision of novel open-source Electronic Design Automation (EDA) tools for reliability assessment of 3D systems, integrating advanced algorithms for physical- and system-level reliability analysis.
\end{abstract}


\section{Introduction}
For decades, semiconductor scaling has been driven by transistor miniaturization, following Moore’s Law and Dennard scaling principles. However, as process nodes enter the angstrom era, further transistor shrinkage faces fundamental physical barriers. The transition from FinFET to gate-all-around (GAAFET) and ultimately complementary FET (CFET) technologies is pushing silicon integration toward its physical limits. As a result, the next leap in performance and energy efficiency will no longer come from planar scaling alone, but from 3D heterogeneous approaches. Key alternatives include fine-grained 3D integration enabled by wafer-to-wafer hybrid bonding \cite{brunion2025cmos}, as well as chiplet-based design strategies.
Three-dimensional chiplet-based architectures provide a practical approach to overcoming the main challenges of modern semiconductor technologies. By partitioning large monolithic dies into smaller, higher-yield chiplets and integrating logic, memory, and accelerators fabricated in different technology nodes, 3D integration enables modular, high-performance, and cost-efficient systems \cite{3doverview}. 


Emerging heterogeneous 3D integration approaches provide a powerful expanded design space, but they also introduce critical challenges related to power delivery, heat removal and thermal management, interconnect reliability, manufacturability, and design automation \cite{3dchallenges}. Addressing these challenges requires a fundamental rethinking of the design ecosystem. Although some dedicated Electronic Design Automation (EDA) tools for 3D integration exist, their capabilities and scalability remain limited compared to mature 2D design environments. As a result, numerous pseudo-3D flows have emerged that rely on adapting commercial 2D tools to approximate 3D design processes \cite{7827574}. However, open-source alternatives remain scarce, restricting access and experimentation for both academia and industry. This lack of open frameworks restricts progress in architectural exploration, reliability analysis, and cross-layer optimization, which are essential to advance next-generation heterogeneous systems.

To this end, the COIN-3D (Collaborative Innovation in 3D VLSI Reliability) aims to strengthen research excellence in this domain. The project aims to develop open-source EDA reliability tools tailored for 2.5D and 3D VLSI designs, while fostering collaboration, mobility, and training among European research institutions. More specifically, the primary objectives of this project are to:
\begin{itemize}
    \item Develop advanced methodologies and algorithms for reliability-aware design and analysis of 3D integrated systems, covering multiple abstraction levels from device to system.
    \item Create a suite of open-source EDA tools enabling multi-physics modeling of key degradation mechanisms, such as power integrity, thermal stress, electromigration (EM), and aging.
    \item Promote training, knowledge transfer, and researcher mobility to strengthen human capital and build long-term expertise in reliability modeling and open-source EDA development in the EU region.
\end{itemize}
Finally, COIN-3D contributes to the objective of the European Chips Act of reinforcing Europe’s technological autonomy and design sovereignty \cite{EUChipsAct2023}. By combining open science practices, open-source EDA development, and AI-aware techniques, the project lays the foundation for a new generation of accessible, intelligent, and sustainable design methodologies.
\begin{figure*}[!hbt]
    \begin{center}
        \includegraphics[width=1.9\columnwidth]{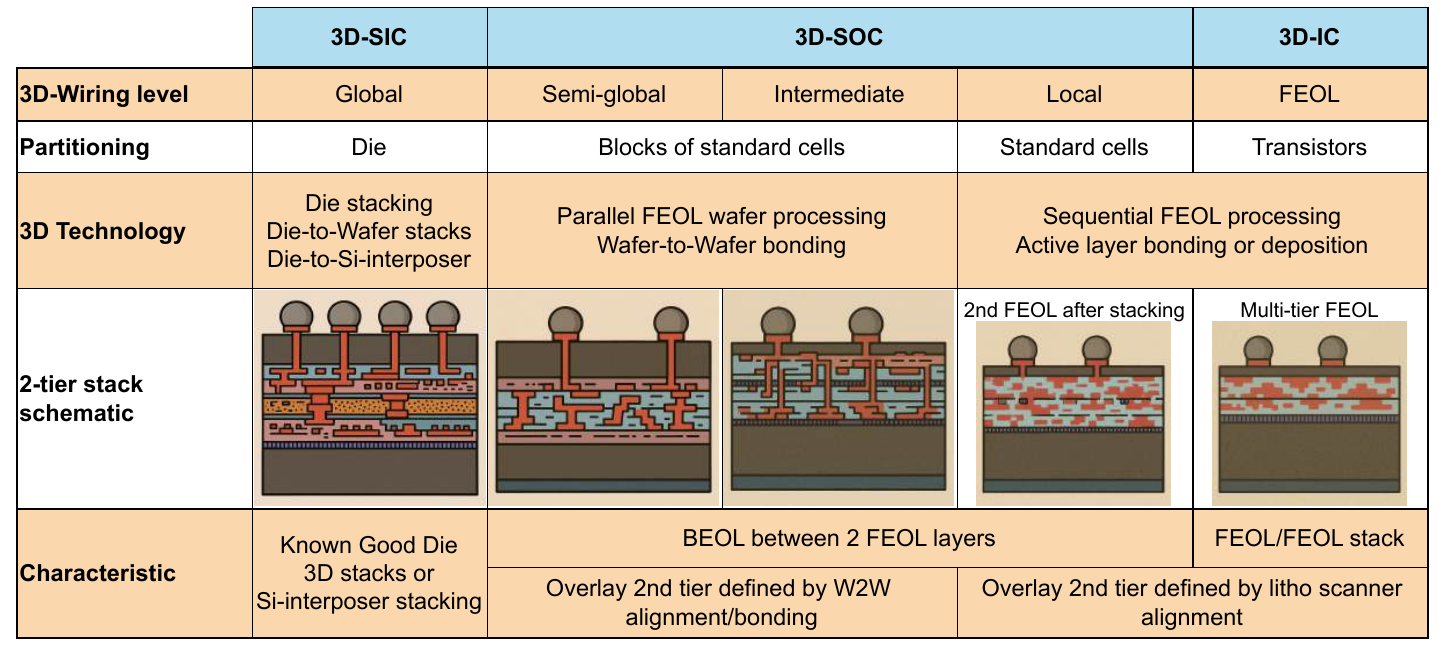}
        \caption{3D integration paradigms. Adapted from \cite{3dlandscape-Trans16}.}
        \label{fig: 3D Integration modes layer stacks}
    \end{center}
\end{figure*}

The remainder of this paper is organized as follows. Section \ref{rel} provides background information and summarizes related work on advanced integration technologies, reliability modeling, and open-source EDA development for 3D VLSI systems. Section \ref{coin} presents the concept of the COIN-3D project, including its scientific objectives, methodological approach, and planned joint research and training activities. Finally, Section \ref{conl} concludes the paper and outlines the main expected outcomes and long-term impact of the project.

\section{Background and Related Work}\label{rel}
Modern VLSI design relies on a well-established set of
physical design practices and tools. However, as device scaling
approaches fundamental limits, 3D integration has emerged as
a disruptive approach that stacks multiple dies vertically to
improve performance, power, and area. This shift in integration
strategy introduces new physical design challenges that require
extending traditional methodologies to handle multi-tiered
designs and complex inter-die connections.
\subsection{3D Integration paradigms}
3D integration offers a broad and evolving range of possibilities, depending on the targeted wiring and partitioning level --such as 3D-SIC, 3D-SoC, and 3D-IC-- as well as the specific technology employed (e.g., die stacking, parallel FEOL wafer processing, wafer-to-wafer bonding, sequential FEOL, etc.). Figure~\ref{fig: 3D Integration modes layer stacks}, adapted from the classical IMEC taxonomy \cite{3dlandscape-Trans16}, summarizes these alternatives, which we briefly outline below.
         
\subsubsection{Fabrication Approach \cite{kim2023seamlessmonolithicthreedimensionalintegration}, \cite{electronics11193013}}
From a fabrication perspective, 3D integration can be classified into monolithic (sequential) and parallel (non-sequential) approaches. Monolithic integration involves fabricating multiple device layers sequentially on the same wafer, resulting in very fine-grained vertical interconnects but posing significant process complexity. Non-sequential  integration, in contrast, combines separate dies or wafers that are fabricated separately using different technology nodes or PDKs (e.g. stacking a 45nm logic layer atop a 7nm tier), enabling functional diversity and technology specialization.

\subsubsection{Stacking paradigms \cite{1008295}}
Three main integration paradigms are commonly used for stacking multiple dies/wafers in 3D ICs. These are F2F, F2B, and back-to-back (B2B). In F2F stacking, the active layers of two dies face each other, enabling very dense interconnects with short signal paths and reduced parasitics, especially when modern wafer-to-wafer hybrid bonding is used. The F2B approach stacks the active side of one die onto the back side of another, which requires the use of through-silicon vias (TSVs) and introduces increased routing complexity. In B2B stacking, the back sides of both dies are placed together, requiring TSVs through both dies. This method is typically used when interconnect density is less critical or when specific thermal and mechanical constraints apply.

\subsubsection{Functional Layering}
3D integration also varies based on the functional composition of the stacked layers. The two primary approaches are memory-on-logic \cite{article}, in which memory dies are stacked above logic dies to enhance bandwidth and reduce latency, and logic-on-logic, where multiple logic layers are stacked to increase computational density. Hybrid architectures that combine these approaches are often preferred, as they offer a balance between performance and area optimization tailored to specific application requirements.

\subsection{Related Work}

\subsubsection{Physical Design Methodologies}
Despite significant advances in 3D integration technologies, the current state of 3D-IC design methodologies still lacks a comprehensive and unified framework for architectural exploration. Designers are often required to make critical implementation decisions, such as die partitioning, stacking order, and inter-tier interconnect topology, very early in the design cycle, often without sufficient analytical feedback. This premature commitment substantially increases design cost and risk. Industrial EDA tools typically rely on proprietary infrastructures that enforce fixed partitioning schemes, while most academic approaches address only isolated design stages such as placement or routing without offering full-flow integration \cite{5621044}. Furthermore, many academic tools focus exclusively on flat or single-die implementations, lacking the hierarchical design capabilities necessary for modern, complex SoCs \cite{markov2012}. Even open-source alternatives such as Open3DBench \cite{3dbench} impose restrictive assumptions by limiting tier configurations to memory-on-logic stacks, while more general frameworks such as Hier-3D \cite{hier3d} still depend on commercial toolchains. Consequently, there remains a lack of open, flexible, and hierarchical exploration frameworks that enable systematic design-space exploration (DSE) for 3D-IC architectures. This limitation forces designers either to over-provision resources, leading to unnecessary area and power overheads, or to risk costly redesign cycles when performance, thermal, or reliability constraints emerge late in development \cite{challenges}.

\subsubsection{Architectural Simulators}

The reliability of modern processors is increasingly shaped by dynamic workload-dependent behavior, which causes significant fluctuations in power consumption, temperature, and voltage over time \cite{11262589}. Early architecture-level reliability models recognized these variations and attempted to capture their effects on component aging and failure risk. However, most of these early approaches targeted uni-core systems, where frameworks such as RAMP modeled runtime-dependent degradation and failure probabilities under varying workload conditions \cite{ramp}. Although these models established foundational principles, they are not scalable to contemporary complex many-core and heterogeneous processors, where tight coupling among thermal, power, and performance domains creates highly nonlinear and spatially variable reliability behavior.

In many-core systems, reliability degradation results from multiple interacting factors, including thermal coupling, dynamic power management, workload heterogeneity, and transient voltage fluctuations, none of which can be accurately represented by a single static model. To support thermal-aware design and optimization, fast and accurate simulation frameworks have become essential. Tools such as CoMeT \cite{comet} and 3D-ICE \cite{3dice} estimate transient temperature distributions using power traces and floorplans as inputs, while power models generated by tools such as McPAT \cite{mcpat} interface with architectural simulators like Sniper \cite{sniper} or gem5 \cite{gem5} to capture power-performance dynamics. Sniper provides scalable interval-based simulation, whereas gem5 offers detailed, cycle-accurate modeling at the expense of runtime. Although these simulators enable thermal analysis, most rely on trace-based methods that do not feed temperature data back into performance simulation, limiting their ability to represent runtime interactions between performance and thermals.

\section{COIN-3D Approach}\label{coin}
The COIN-3D project is a Horizon Europe Twinning project that integrates advanced technical research with structured collaboration and knowledge exchange among European academic partners to address the challenges of reliability in 2.5D/3D chiplet-based systems. In addition to its core scientific objectives, the project places strong emphasis on strengthening collaboration between the University of Thessaly and leading European research institutions, including the University of Amsterdam, the University of Bremen, and the University of Glasgow. This collaborative framework is designed to promote sustained interaction, mutual learning, and the dissemination of best practices across institutions with complementary expertise.

A central element of this approach is the systematic organization of training and knowledge-sharing activities that complement the technical research. These activities include scientific webinars, advanced technical courses, focused workshops, and dedicated summer schools addressing key topics such as 3D integration technologies, reliability modeling, electronic design automation, and emerging chiplet-based architectures. Through these initiatives, COIN-3D supports skills development for researchers and early-career scientists, while fostering a shared understanding of state-of-the-art methodologies, tools, and design practices within the consortium.

\begin{figure}[!hbt]
    \begin{center} 	
    \includegraphics[width=1\linewidth,keepaspectratio]{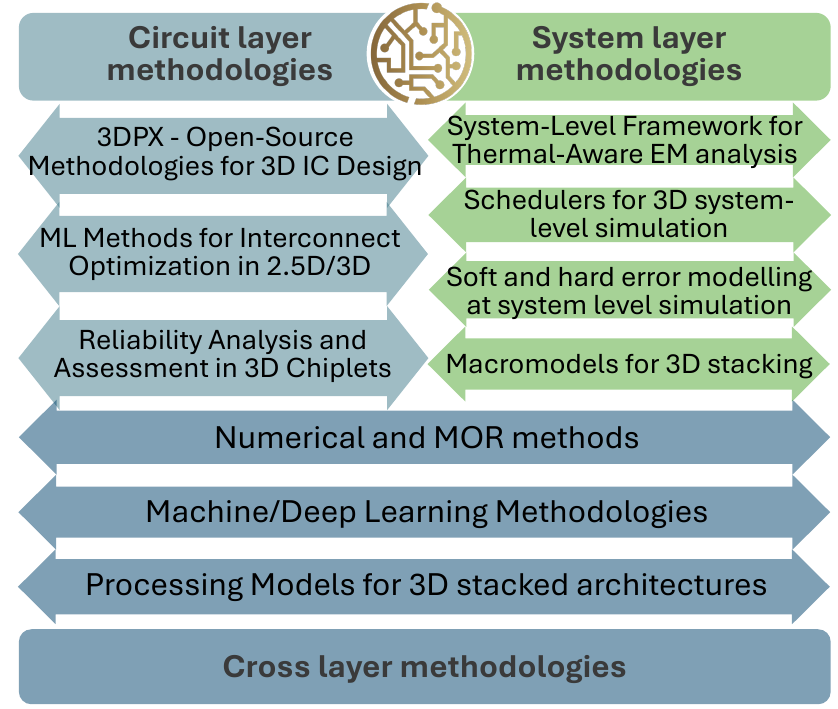}
    \end{center}
    \caption{COIN-3D methodologies across different abstraction layers.}
    \label{fig:methods}
\end{figure}

COIN-3D builds on the accumulated expertise of the research teams in 3D design flows, thermal analysis, and reliability modeling, as well as on the experience gained from tools developed by the participating teams over the past few years, such as \cite{hier3d, comet, proton}. As a result, the research component of COIN-3D seeks to develop a comprehensive tool set that bridges the gap from circuit-level to system-level analysis, addressing the critical challenges of 2.5D/3D chiplet architectures, as shown in Fig. \ref{fig:methods}. The following subsections describe the main technical methodologies and research directions pursued within COIN-3D across the physical and system abstraction layers.

\subsection{Physical level methodologies}

\noindent\textbf{3D Physical Design Exploration}\\
A central scientific objective of COIN-3D is to advance the state of physical design automation for heterogeneous 3D integration. The project focuses on developing methodologies and open-source tools that enable accurate modeling and optimization of physical-level effects, such as power delivery, thermal coupling, and reliability, across stacked integrated circuit tiers. Within this context, COIN-3D has produced 3DPX\footnote{https://github.com/coin3d-project/coin3d-project/tree/main/3DPX\_testcase} (An Open-Source Methodology for 3D Physical Design Exploration), the project’s principal physical-level research output \cite{3dpx}.

3DPX provides the first fully open-source framework for hierarchical 3D design exploration built upon the OpenROAD infrastructure \cite{openroad}. It extends traditional two-dimensional design flows to support a wide range of integration paradigms, including F2F, F2B, and monolithic stacking. Through a modular and hierarchical flow, 3DPX allows designers to investigate logic-on-logic, memory-on-logic, and hybrid architectures with multi-PDK compatibility. This flexibility enables exploration across different technology nodes and bonding schemes, making it possible to evaluate physical-level trade-offs early in the design process without dependence on proprietary commercial tools.

At its core, 3DPX introduces a hierarchical 3D place-and-route methodology that manages design partitioning, floorplanning, placement, and routing across multiple dies. The flow extends standard ASIC collaterals, LEF, DEF, and LIB, in order to incorporate 3D tier information, inter-tier vias, and routing layers. A dedicated Python- and TCL-based API automates the modification of technology files and manages the synchronization of design data across tiers. This approach maintains consistency with existing OpenROAD data structures while enabling concurrent implementation of multi-tier blocks, reducing design turnaround time, and enhancing scalability for large-scale system-on-chip SoC architectures.\\

\noindent\textbf{Electromigration-Aware Physical Reliability Modeling}\\
A key element of COIN-3D’s physical-level research focuses on developing methodologies for electromigration (EM) modeling and reliability-aware optimization, essential for ensuring long-term integrity in 3D VLSI architectures \cite{chen2017analytical}. As interconnect dimensions shrink and current densities rise, EM has emerged as one of the most critical degradation mechanisms, directly affecting the lifetime and stability of power delivery networks and signal interconnects \cite{lienig}. Within COIN-3D, EM analysis is approached from both physics-based and data-driven perspectives, combining accuracy with computational efficiency and early design applicability.

At the physics level, COIN-3D builds upon advanced formulations of Korhonen’s diffusion model, which describes the evolution of hydrostatic stress under high current density and temperature gradients \cite{Korhonen}. Using this model as a foundation, the project developed an open optimization framework that employs numerical solvers and gradient-based optimization to evaluate and mitigate EM-induced stress \cite{axelou2025towards}. This approach enables design-space exploration where interconnect geometry, material parameters, and layout constraints can be adjusted to minimize stress accumulation and IR drop while maintaining design rule compliance. The resulting optimization flow has demonstrated a significant improvement in stress uniformity and power integrity for large-scale power grid benchmarks, showing great potential for integration into the open 3DPX environment.

To complement these physics-driven techniques, COIN-3D integrates machine learning–assisted EM prediction to accelerate early-stage reliability analysis \cite{11083905}. Using a compact feature set, including current density, diffusivity, spatial location within interconnect segments, and temperature, the model employs the Extreme Gradient Boosting (XGBoost) algorithm \cite{chen2016xgboost} to predict EM stress with high accuracy based on data derived from detailed physics simulations. This allows designers to estimate potential EM hotspots at an early phase of the design flow, before full routing and parasitic extraction are available. 

Together, these methodologies are used for the physical-level reliability framework within COIN-3D. The hybrid approach merges interpretability and precision from physical modeling with the scalability and speed of machine learning.  By embedding reliability evaluation directly into the physical implementation stage, COIN-3D aims to enable  early detection and mitigation of degradation mechanisms, reducing design iterations and improving the predictability of advanced 3D systems.

\subsection{System level methodologies}
\noindent\textbf{System-Level Aware PDN Generator}\\
The first major component of COIN-3D’s system-level research focuses on developing a system-aware power delivery network (PDN) generation and analysis methodology \cite{pantazi2025system}. Traditional PDN design flows often rely on static current profiles or uniform grid templates that fail to represent the highly dynamic and spatially non-uniform power consumption observed in modern heterogeneous and 3D processors. To overcome these limitations, COIN-3D implements a die-centric PDN generation framework that leverages workload-dependent power traces derived from architectural simulators such as \cite{8444047}.

The proposed PDN synthesis tool constructs resistive meshes that reflect this behavior, optimizing metal-layer dimensions, grid density, and pad allocation to minimize IR drop while maintaining EM safety margins. This methodology establishes an early link between architectural behavior and physical PDN reliability, allowing designers to explore multiple PDN configurations before layout implementation. The resulting workflow enables predictive EM and IR-drop evaluation under realistic workload conditions, significantly reducing design iteration time and enhancing confidence in early-stage design decisions.\\
\\
\noindent\textbf{System Level Thermal-Aware Reliability Analysis}\\
Although physical-level design determines the intrinsic robustness of interconnects and devices, system-level behavior ultimately defines how these components are stressed under real workloads. In contemporary heterogeneous architectures, reliability degradation emerges from dynamic interactions between computation, power management, and thermal control, which evolve over time and across spatial hierarchies. The COIN-3D project therefore extends its focus beyond the physical implementation stage to establish a system-level reliability framework that connects device-level degradation mechanisms with architectural behavior and workload-driven stress.

COIN-3D project adopts a cross-layer modeling approach that bridges the gap between physics-based reliability models (e.g., EM and soft-error rate) and system-level operational parameters (e.g., power density, thermal gradients, and workload distribution). Using hierarchical abstraction, lower-level degradation models are expressed as compact behavioral equations that can be embedded into architectural simulators. This enables runtime estimation of lifetime impact for different voltage-frequency settings, scheduling policies, and cooling strategies.

A key feature of this methodology is its ability to perform bidirectional data exchange between the architectural simulators and the reliability engine. Temperature and current information generated by performance or thermal simulators, such as HotSniper \cite{8444047}, is fed into the reliability models, while the resulting degradation metrics, such as EM degradation and aging,  feed back into the architectural loop. This two-way coupling allows system-level decisions, like task migration or dynamic voltage scaling, to be evaluated in terms of both performance and long-term reliability, supporting design-time exploration and runtime control within the same framework.

\section{Conclusions}\label{conl}


Since semiconductor technology has moved beyond the traditional limits of planar scaling, fine-grained heterogeneous 2.5D and 3D integration has emerged as a key enabler for continued advances in computing performance, energy efficiency, and system heterogeneity. While these integration paradigms offer significant opportunities, they also introduce critical challenges related to reliability, testability, and design automation, which must be addressed in a systematic and scalable manner.

In this context, the COIN-3D Horizon Europe Twinning project establishes a collaborative framework for advancing reliability-aware design methodologies for 2.5D and 3D VLSI systems. Through the coordinated efforts of its European research partners, the project develops open-source EDA tools and methodologies capable of modeling and analyzing reliability aspects across multiple abstraction levels, from physical implementation to system-level behavior. By combining physical modeling, system-level analysis, and cross-layer design approaches, COIN-3D contributes to strengthening the scientific foundations required for reliable heterogeneous integration.

The expected outcomes of the project include the development of advanced reliability-aware design tools, enhanced capabilities for early-stage design-space exploration, and the establishment of sustainable open-source frameworks that support continued innovation beyond the project’s duration. In the longer term, COIN-3D aims to foster cross-disciplinary collaboration, stimulate new research directions in 3D reliability modeling and design automation, and contribute to strengthening Europe’s capacity and competitiveness in advanced semiconductor research and development.

\section*{Acknowledgment}
This work was funded under the COIN-3D project, which has received funding from the European Union’s Horizon Europe research and innovation program under grant agreement No. 101159667. The content reflects only the authors’ views, and the European Commission is not responsible for any use that may be made of the information it contains.

\balance
\bibliographystyle{IEEEtran}
\bibliography{Main}
\end{document}